\documentclass[lettersize,journal]{IEEEtran}
\usepackage{amsmath,amsfonts}
\usepackage{algorithmic}
\usepackage{algorithm}
\usepackage{array}
\usepackage[caption=false,font=normalsize,labelfont=sf,textfont=sf]{subfig}
\usepackage{textcomp}
\usepackage{stfloats}
\usepackage{url}
\usepackage{multirow}
\usepackage{verbatim}
\usepackage{graphicx}
\usepackage{cite}
\begin{document}

\title{Fibonacci-Net: A Lightweight CNN model for Automatic Brain Tumor Classification}

\author{Santanu Roy, Ashvath Suresh, Archit Gupta, Shubhi Tiwari, Palak Sahu, Prashant Adhikari, Yuvraj S. Shekhawat
        % <-this % stops a space
\thanks{Santanu Roy, a former employer of Christ (Deemed to be University), is now working at NIIT University, Rajasthan, India (Email: santanuroy35@gmail.com). 

Ashvath Suresh and Prashant Adhikari are working with the CSE department, Christ (Deemed to be University), Bangalore, India. (Email: ashvath.suresh@btech.christuniversity.in; prashantprofessional99@gmail.com). 

Whereas, Archit Gupta, Shubhi Tiwari, Palak Sahu, and Yuvraj S. Shekhawat are working with the Computer Science and Engineering Department, NIIT University, Jaipur, India.

(Email: gupta.archit2001@gmail.com; tshubhi2807@gmail.com; palak.sahu20@st.niituniversity.in, yuvrajsingh.shekhawat1310@gmail.com).
}% <-this % stops a space
\thanks{Manuscript received Jan 24, 2025; revised xxxxx.}}

\markboth{IEEE Transactions on xxxx}%
{Shell \MakeLowercase{\textit{et al.}}: A Sample Article Using IEEEtran.cls for IEEE Journals}

\IEEEpubid{}

\maketitle

\begin{abstract}
This research proposes a very lightweight model ``Fibonacci-Net'' along with a novel pooling technique, for automatic brain tumor classification from imbalanced Magnetic Resonance Imaging (MRI) datasets. Automatic brain tumor detection from MRI dataset has garnered significant attention in the research community, since the inception of Convolutional Neural Network (CNN) models. The potential impact of such advancements in healthcare is profound, and it has the potential to improve the accuracy of diagnosis. However, the performance of conventional CNN models is hindered due to class imbalance problems, since most of the medical datasets are found imbalanced in nature. The novelties of this work are as follows: (I) A lightweight CNN model is proposed in which the number of filters in different convolutional layers is chosen according to the numbers of Fibonacci series. (II) In the last two blocks of the proposed model, depth-wise separable convolution (DWSC) layers are employed to considerably reduce the computational complexity of the model. (III) Two parallel concatenations (or, skip connections) are deployed from $2^{nd}$ to $4^{th}$, and $3^{rd}$ to $5^{th}$ convolutional block in the proposed Fibonacci-Net. This skip connection encompasses a novel Average-2Max pooling layer that produces two stacks of convoluted output, having a bit different statistics. Therefore, this parallel concatenation block works as an efficient feature augmenter inside the model, thus, automatically alleviating the class imbalance problem to a certain extent. For validity purpose, we have implemented the proposed framework on three MRI datasets which are highly class-imbalanced. (a) The first dataset has four classes, i.e., glioma tumor, meningioma tumor, pituitary tumor, and no-tumor. (b) Second and third MRI datasets have 15 and 44 classes respectively. Experimental results reveal that, after employing the proposed Fibonacci-Net we have achieved 96.2\% accuracy, 97.17\% precision, 95.9\% recall, 96.5\% F1 score, and 99.9\% specificity on the most challenging ``44-classes MRI dataset''. This is so far the best result on this dataset. Furthermore, the proposed framework also generalized well on other two MRI datasets. All the codes of several experiments along with their graphs, and confusion matrices are available on a GitHub link:

{(https://anonymous.4open.science/r/Brain-Tumor-Detection/)}
\end{abstract}


\section{Introduction}

\IEEEPARstart{H}uman brain is one of the most vital organs in the body, acting as the central hub for regulation and coordination [1]. A brain tumor refers to the abnormal proliferation of cells within the brain (or, the central spinal cord), which interferes with the nervous system and can be life-threatening [2]. Brain tumors can be categorized mainly into three types, gliomas, pituitary, and meningiomas tumors [3]. Effective and appropriate treatment of brain tumors is only possible if they are detected at an early stage. One of the most reliable methods for doing this is through Magnetic Resonance Imaging (MRI) scans of the brain [4]. MRI is a noninvasive, painless diagnostic imaging technique that provides detailed images of internal body organs in both 2D and 3D formats. Moreover, unlike X-rays, MRI does not require radiation exposure. Neurologist experts then examine these MRI scans to detect abnormalities, particularly in the brain’s lining, that are caused by tumors. However, manual detection of brain tumors from MRI scans remains highly complex, labor-intensive, and time-consuming process. Merely identifying the presence of a brain tumor is not sufficient for doctors to initiate treatment. It is crucial to accurately classify the type of tumor [3], as a miss-classification could have life-threatening consequences for the patient, making the process more time-consuming. Hence, there is an urgent requirement of Computer Assisted Diagnosis (CAD) for brain tumor classification from MRI images [5].

Recent advancements in Deep Learning have demonstrated enhanced performance for the automatic brain tumor detection. In particular, Convolutional Neural Networks (CNNs) have been extensively utilized by numerous researchers [6-9]. Hossam H. Sultan et al. [10] developed a CAD system that begins by performing data augmentation through techniques such as flipping, rotation, and furthermore adding salt-and-pepper noise to the images. The system then classifies MRI images into different tumor types, that are, gliomas, pituitary, and meningiomas tumors, using a CNN model. Ahmed et al. [11] introduced a Deep Convolutional Neural Network (DCNN) architecture for 4 types of tumor classification. Their methodology involved a three-step pre-processing method, which included removing irrelevant elements, de-noising using the Non-Local Means (NLM) algorithm, and contrast enhancement by histogram equalization. Their approach achieved an average accuracy of 98.22\% on a single dataset. Although their approach performed effectively on a specific dataset, there is no guarantee that it would yield similar results on other datasets. M. Rizwan et al. [9] have proposed a Gaussian CNN in order to classify several grades of tumor. They have utilized Gaussian filter in order to augment the small data. Despite having higher efficacy, employing extensive use of pre-processing method (Gaussian filter) is not so feasible, which may lead to potential data loss. In a separate study, Sarah Ali Abdelaziz Ismael [7] implemented a Residual Neural Network (ResNet) for diagnosing brain cancer using MRI scans. The proposed method utilized a Residual CNN model, alongside various image augmentation techniques, achieving an accuracy of 99\% at the image level and 97\% at the patient level. Likewise, Deepak et al. [6] created a CNN-SVM model in which CNN was utilized for only the feature extraction task. The final classification is accomplished by SVM. Their proposed framework had an accuracy of 97.8\% ± 0.2\% in differentiating gliomas, meningiomas, and pituitary tumors.

Abiwinanda et al. [12] reported that without employing data augmentation or pre-processing techniques, the performance of a basic CNN model is hindered due to class imbalance problem. They explored five different possible configurations of CNN (trained from scratch) on a 4-class MRI dataset. In three of the models, they utilized 1, 2, and 3 numbers of convolutional layers respectively, while in the remaining two models, they varied the number of dense layers. Given the substantial class imbalance in the 4-class dataset, the comparatively heavy model (with 3 convolutional layers) performed marginally better, but it was highly susceptible to significant overfitting. Their most effective CNN configuration could not achieve more than  84.19\% validation accuracy. The aforementioned studies also revealed that there remains a significant research scope in the field of automatic brain tumor detection from MRI scans. The primary challenge lies in developing a novel model (trained from scratch) that can perform efficiently on any imbalanced MRI dataset (for brain tumor detection) in a very generalized manner, such that it can eventually serve as a universal standard model for researchers.

An alternative approach of resolving this class imbalance could be employing cost sensitive Deep Learning model [13]. For instance, instead of utilizing the Categorical Cross Entropy (CCE) loss function, one might opt for a weighted CCE (WCCE) loss function or, Balanced WCCE (BWCCE) [14] where the weights assigned to each class are inversely proportional to the number of instances in that class, as well as sum of all such weights is equal to 1 to ensure that it satisfies the notion of probability. Nevertheless, M. Tyagi et al. [15] have addressed that WCCE or BWCCE might introduce some fluctuations in the performance of the dominant class, if the dataset is severely imbalanced. S. Deepak et al. [16] have proposed a weighted focal loss within a CNN framework to mitigate the class imbalance issue. However, the authors asserted that determining these class weights is a challenging endeavor, and even minor adjustments to those weights could lead to significant fluctuations in the model performance.

Another approach to this problem could be employing recent trend transformer models such as Vision Transformer (ViT) [17], Swin transformer [18]. Despite being an analogous model to the transformer (used extensively in NLP), the Vision Transformer is not yet the default choice for researchers in computer vision [19]. Its higher complexity makes it incompatible for small datasets, moreover, it lacks the multi-scale hierarchical framework [20] (found in CNNs), which is crucial in the computer vision task. Due to the above mentioned reasons, numerous researchers have integrated both the notions of ViT and CNN. Recently, Byeongho Heo et al. [21] have proposed a Pooling-based ViT (PiT) that integrates pooling layers into the ViT model. P.T. Krishnan et al. [22] have proposed a rotation invariant Vision Transformer (ViT) in order to detect binary classes: (I) glioma tumor, and (II) no tumor. Authors claimed that incorporating `rotation invariance' can enhance the robustness and generalization capabilities of the transformer model. Ishak Pacal1 [23] has proposed a Swin transformer for classifying MRI images into 4 classes. Moreover, they have incorporated Residual MLP, instead of MLP in their model that enhanced the efficacy by 1-2\%.

There is another way of leveraging attention mechanism to the deep learning model, that is, incorporating a channel attention or spatial attention directly into a CNN model. J. Hu et al. [24] first time proposed a Squeeze Excitation Network (SE-Net) which is a notable breakthrough in the field of computer vision, known for providing channel attention. S.Woo et al. [25] invented Convolutional Block Attention Module (CBAM) in which they integrated both channel attention and spatial attention in order to enhance the generalization ability of the model. M. Togacar et al. [26] have proposed a novel BrainMRNet that consists of 3 distinct convolutional blocks. In $2^{nd}$ and $3^{rd}$ block they employed CBAM attention blocks, moreover, they have concatenated the output from $1^{st}$ block and $2^{nd}$ block to the GAP layer by utilizing an up-sampling mechanism. Despite achieving decent efficacy, their method may lead to significant loss of information due to heavy usage of interpolation mechanism. A.B.Abdusalomov et al. [27] have proposed a pre-trained Yolo-v7 model to accurately detect meningioma, glioma, and pituitary gland tumors from MRI scans. Furthermore, they have incorporated CBAM attention module on Yolo-v7 model in order to further enhance the model performance. AG Balamurugan et al. [28] utilized a ResNet-101 combined with a Channel-wise Attention Module (CWAM) to effectively classify brain tumor grades using MRI scans. They noted a substantial improvement in performance after integrating the CWAM block alongside the pre-trained CNN model. Many more relevant research can be found in [29-35]. 

\subsection{Contributions of the paper}
The contributions of this paper are as follows: 
\begin{enumerate}
        \item A very lightweight CNN model, Fibonacci-Net, is proposed where the number of filters of several convolutional layers is chosen according to the Fibonacci Series.
	\item Two parallel concatenation blocks (pcb) or skip connections are incorporated in the proposed framework. In these pcb, novel pooling technique ``Avg-2Max Pooling'' are leveraged for the first time. 
 \item This ``Avg-2Max Pooling'' converts the image into kind of its negative image, where edges get extra attention. This pcb works as an automatic feature augmenter inside the model, thus, the proposed framework automatically mitigates the class imbalance problem to some extent. 
 \item Experimental results reveal that the proposed ``Fibonacci Net'' along with two pcbs, has superior performance than the recent trends models Pooling based Vision Transformer (PiT), ConvNext-V2 etc., and other state-of-the-art models, as well as it generalizes well across 3 diverse MRI datasets. 
 \item Furthermore, we have validated our proposed theory by ``Explainable AI". A GradCam heat map diagram that showed that the proposed ``Fibonacci Net'' with pcb, provides more attention to tumor regions compared to the same without pcb.
\end{enumerate}

\section{Dataset and Challenges}
We have implemented the proposed framework on three MRI datasets for checking validity of the proposed framework. (I) The first MRI dataset encompasses four classes [36], that are, glioma tumor, meningioma tumor, pituitary tumor, and no-tumor. The number of images in the glioma tumor, meningioma tumor, pituitary tumor, and no-tumor class are 926, 937, 901, and 500 respectively. Hence, it is not a severe class imbalance problem. (II) The second MRI dataset [37] is comparatively more imbalanced dataset than the previous one, containing a total of 4479 images. These images are acquired by the collection of different modes of MRI scans, including T1, T2, and contrast-enhanced T1 (T1C+). Total 15 classes are present in this dataset which are astrocytoma, carcinoma, ependymoma, ganglioglioma, germinoma, glioblastoma, granuloma, medulloblastoma, meningioma, neurocytoma, oligodendroglioma, papilloma, schwannoma, tuberculoma and normal class. Additionally, each category is further divided into three types of scan modes: T1, T1C+, and T2. However, the normal class includes only two modes, T1 and T2. By treating each scan mode as a separate category, this dataset can is further expanded into a 44-class MRI dataset. This 44-class dataset is severely imbalanced, with some classes having very few samples. For instance, the ``Granuloma T2'' class contains just 17 images, while the ``Tuberculoma T1'' class has only 28 images. On the other hand, the largest category, ``Meningioma T2'', includes 233 images. The combination of 44 distinct classes and severe class imbalance makes this dataset particularly challenging. As a consequence, many existing CNN models have struggled to perform effectively on this MRI dataset, particularly on minor classes. 
\begin{figure*}[h]
		\centering
		\includegraphics[width=17.8cm,height=6.3cm]{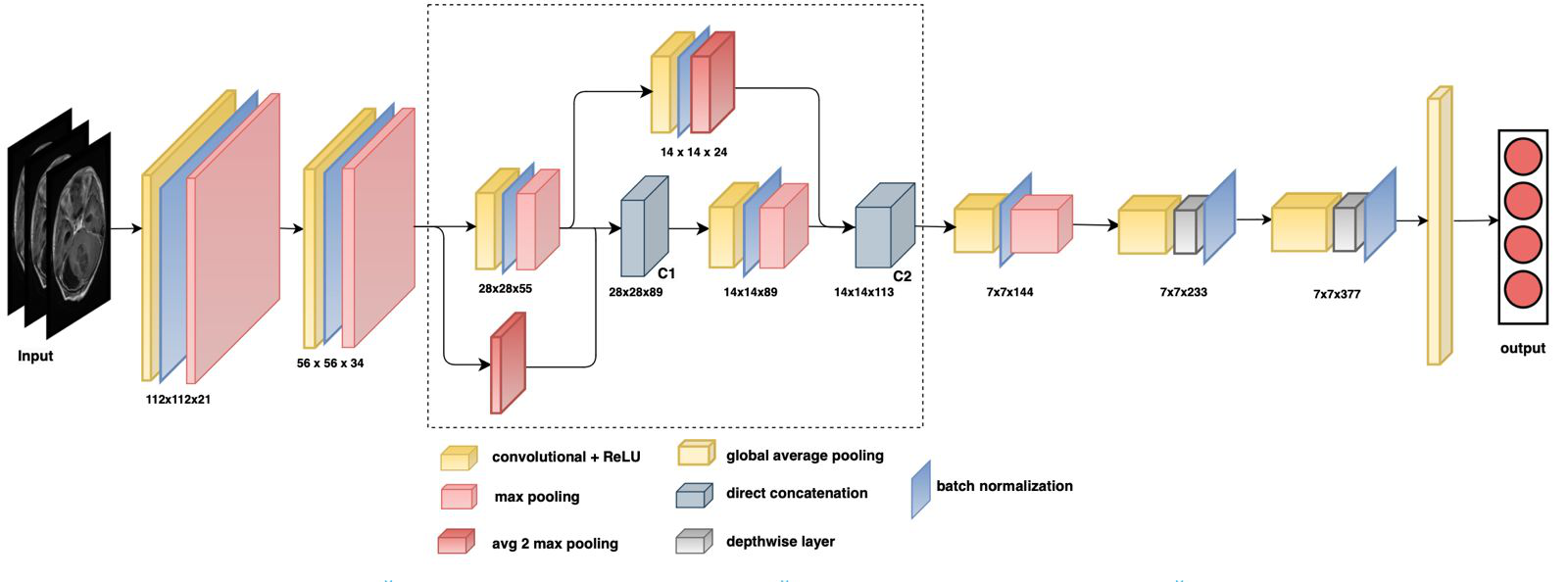}
		\caption{Block Diagram of the proposed Fibonacci Net}
	\end{figure*}
 \begin{figure*}[h]
		\centering
\includegraphics[width=12.8cm,height=6.0cm]{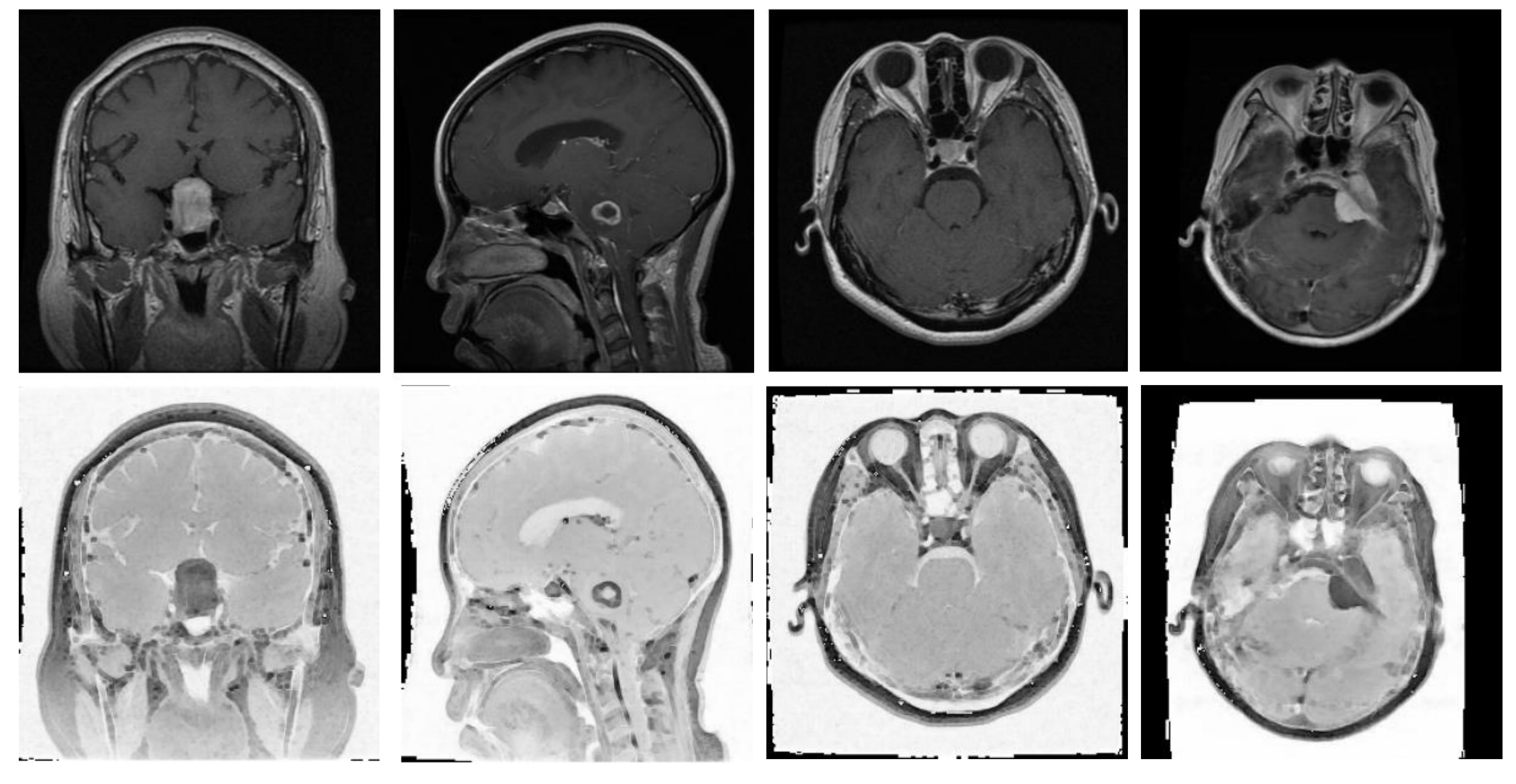}
		\caption{The first row indicates original MRI images; the second row are the down-sampled (/2) images after passing it through Avg-2Max pooling }
	\end{figure*}
\section{Methodology}
The proposed ``Fibonacci-Net" consists of two basic units: One is conventional (sequential) block, the other is parallel concatenation block (pcb). The entire proposed methodology is further explained in depth in four following subsections: (a) Fibonacci-Net Architecture, (b) Avg-2Max Pooling technique, and the parallel concatenation, (c) Mathematical analysis of pcb, (d) Training specifications. 

\subsection{Fibonacci-Net Architecture}
A lightweight CNN model architecture `Fibonacci-Net' is first time proposed in this research for brain tumor detection from MRI images, which is shown in Fig.1. In this `Fibonacci-Net', the number of filters (in the convolutional layers) is chosen according to the \textit{Fibonacci series} numbers starting from 21, 34, 55, 89, 144, 233 and 377 etc. Unlike other CNN models like VGG-16 [38] or Mobile-Net [39], here the number of filters is not exponentially raised in consecutive convolutional blocks, rather this number of filters is chosen by the sum of two numbers of filters in the previous two Convolutional blocks (i.e., following \textit{Fibonacci series}). Therefore, the number of filters with consecutive convolutional blocks is exceptionally less. As a consequence, the spectral dimension and the overall number of trainable parameters are notably reduced in the proposed model. In this Fibonacci Net, we have managed to limit the number of hyper-parameters in the model to only 13.9 lakhs, despite employing two parallel concatenation blocks. Furthermore, after every convolutional layer we deployed 1 batch normalization layer. This batch normalization layer ensures a smooth gradient flow throughout the network and hence, reduces the over-fitting problem, to a certain extent.

Moreover, one Depth-Wise Separable Convolutional (DWSC) layer [40] is employed in the proposed model at both of its last two blocks. DWSC factorizes the convolution layer into two separate layers, one is Depth-Wise convolution layer which performs filtering operation (by 3$\times$3 convolutional filter) only one time. This is followed by 1$\times$1 convolution layer which does point-wise multiplication, thus, it further reduces the computational complexity considerably from the model.

The number of trainable parameters in the convolutional layer of the proposed Fibonacci Net is
\begin{equation}
m_c=(3\times3\times{f}_0+1).{f}_1 
\end{equation}
Here, in equation (1), ${f}_0$ is the number of filters in the previous layer (or, input layer), and ${f}_1$ is the number of filters in the current convolutional layer. The number of filters is represented by $f$ here, because this is according to \textit{Fibonacci} series numbers. The kernel size in all convolutional layers and DWSC layers is chosen $3\times3$ with stride $1$, keeping zero padding ``same". 

The number of trainable parameters in DWSC layer  of the proposed Fibonacci Net is
\begin{equation}
m_{d}=(3\times3\times1+1).{f}_0=10{f}_0 
\end{equation}
That means the number of trainable parameters in a DWSC layer (with 3$\times$3 kernel) is only 10 times the number of filters in the previous layer.
This can be noticed from equations (1) and (2) that $m_c>>m_{d}$ for deeper layers. Because the number of filters in the last two convolutional blocks are 233 and 377 respectively. Hence, DWSC enables the CNN model to limit its number of hyper-parameters significantly. Furthermore, Global Average Pooling (GAP) layer [39] is deployed instead of flatten layer in the proposed CNN model. This GAP layer takes the average in spatial domain, thus, it only takes the equal number of neurons as the spectral dimension in the last layer (i.e., 377 for the proposed model). On the other hand, the number of neurons for the flatten layer could be 7$\times$7$\times$377. Hence, GAP layer reduces the number of hyper-parameters to a large extent (because the number of neurons is reduced here by 49 times) in the dense layer of the proposed Fibanacci Net. Consequently, the proposed model becomes a lightweight CNN model and thus, avoids over-fitting to a large extent. 

\subsection{Avg-2Max Pooling Technique and Parallel Concatenation}
 Another novelty we bring in the proposed model is the parallel concatenation branch (or, skip connection) which works as a feature augmenter inside the model. This idea of parallel concatenation or skip connection is inspired from Inception-V3 [41] and one recent concatenated AD-Lite Net [42], however, the incorporation of this Avg-2Max Pooling layer in parallel branch is entirely novel. The significance of this skip connection is that, it may improve the gradient flow inside the network as well as it repeats some of the previously extracted features in the model, as a result, CNN model will create more variety of features inside the model now. This parallel concatenation block (pcb) starts from the output of the $2^{nd}$ convolutional block and is concatenated (through a concatenation block) with the main tensor and goes to the input of $4^{th}$ convolutional block, as shown in Fig.1. Another concatenation is deployed (as shown in Fig.1) from the $3^{rd}$ convolutional block to $5^{th}$ convolutional block which encompasses ``Avg-2Max Pooling'' followed by 24 number of 3$\times$3 convolutional filters.

 The proposed ``Avg-2Max pooling'' can be thought of as a function, with one Average-Pooing function and $2$Max-pooling function is subtracted from it. This ``Avg-2Max pooling'' with pool size 3$\times$3 and stride 2, can be represented by the following equation (3).
 
 \begin{equation*}
		{(g_{a2m}{(I_o(f))}_{w\times  w})}_{3\times 3|2}={I_o{(avg(f)}_{3\times 3})}_{\frac{w}{2}\times \frac{w}{2}}-
\end{equation*}
\begin{equation}
		{I_o({max(f)}_{3\times 3}+{max(f)}_{3\times 3})}_{\frac{w}{2}\times \frac{w}{2}}
\end{equation}

The basic idea of this ``Avg-2Max pooling'' is coming from the notion of ``Digital Image Processing (DIP)'' [43]. Here, average pooling with a 3$\times$3 pool size, is analogous to a blur filter (or, average filter) in DIP. On the other hand, the max-pooling layer with a 3$\times$3 pool size, tries to preserve the maximum intensity information inside the convoluted tensor. Thus, it is analogous to the main original image, to some extent. Hence, this ``Avg-2Max pooling'' is equivalent to subtracting the main image from its blurry component. This method has the ability to highlight (or, extract) edges because it is the first difference between a pixel intensity and local mean (intensity) value [44]. Nevertheless, this ``Avg-2Max pooling'' layer not only highlights the edge components, but also, it creates a negative image because it subtracts the original image component twice from its blurry component. Some of the processed images are shown in Fig.2. It is observed from Fig.2, after passing the input tensor through this ``Avg-2Max pooling'' layer, we obtain images in which edge components get extra attention, additionally, it also differs the overall statistics of the image considerably without changing the contextual information much. According to S.Roy et al. [14], any efficient data-augmenter should generate images which have slightly different statistics than original, otherwise, it may cause substantial overfitting, and thus, making the technique less effective. Hence, it can be concluded that proposed ``Avg-2Max pooling'' layer has the capability to generate a distinct set of feature maps (or, augmented images) for the brain tumor detection, thus, the proposed pcb connection can be considered as an efficient automatic feature augmenter, working in the middle of the CNN model. 

\begin{figure*}[h]
		\centering
\includegraphics[width=15.1cm,height=4.2cm]{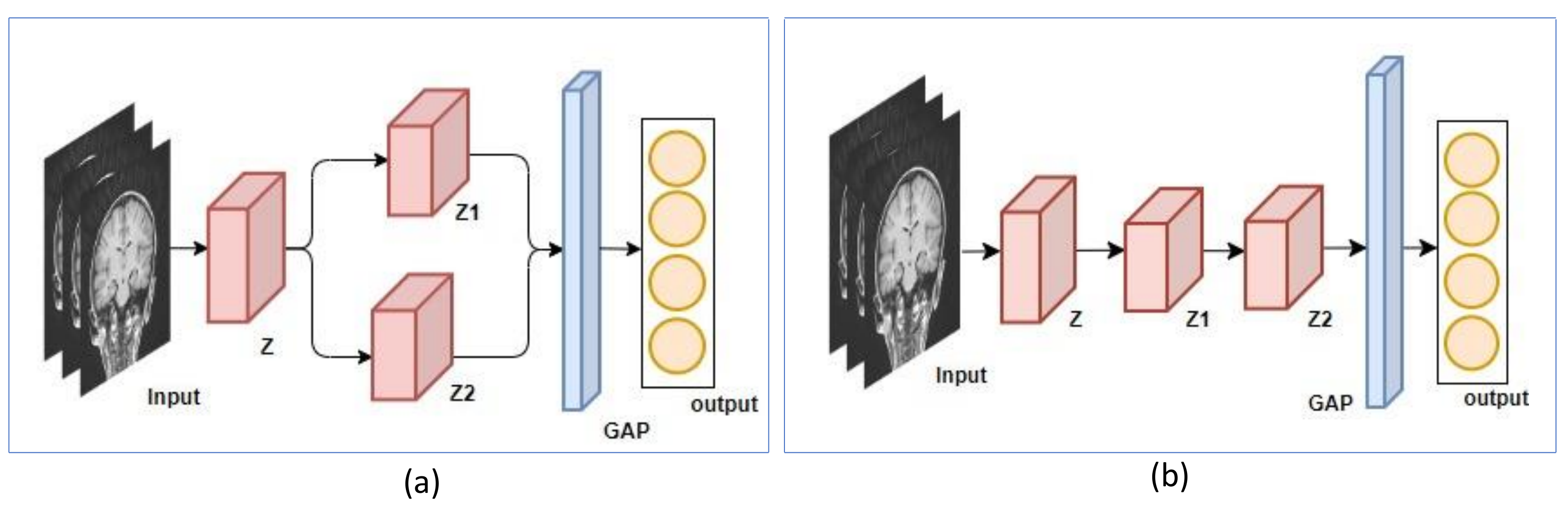}
		\caption{Basic units of the proposed model (a) parallel concatenation vs (b) sequential model}
	\end{figure*}
To the best of our knowledge, this is the first work where this kind of unique pooling technique was leveraged for generating augmented feature maps in the middle of the CNN model. However, one may rise question ``why not utilize this data augmenter at the input of the CNN model?'' The reason is further explained below. 
\begin{enumerate}
\item Some of the previously extracted features (in the previous layers) have been repeated with a bit change of statistics, by incorporating pcb blocks. As CNN model has a tendency to forget earlier features [42], this approach of generating modified features midway through the CNN is highly effective. Because, a diverse set of (previously extracted) features have been mixed with the current layer features, which, in turn, improves the model's generalization capabilities. 
\item We can control the contribution percentage of these augmented features in the proposed CNN model. For example, before the $2^{nd}$ pcb, the spectral dimension of the input tensor was already 55. However, by incorporating 24 numbers of ($3\times 3$) convolutional filters, (prior to $2^{nd}$ ``Avg-2Max pooling'' layer) now that proportion of contribution of augmented features with respect to the main backbone network is only $24:89$. In this fashion, we can reduce or, increase the effect of augmented features in the proposed CNN model, by manipulating the number of filters.
\item Moreover, by employing $3\times3$ convolutional filters preceding the $2^{nd}$ ``Avg-2Max pooling'' layer, we can harness a form of automatic feature extraction operation. Because in this $3\times 3$ convolutional block, the filter coefficients are trainable parameters. Thus, with each epoch, the CNN model endeavors to adjust these coefficients such that it can automatically generate distinctive (augmented) feature maps which are very essential for the brain tumor detection. 
\item Furthermore, relying extensively on data augmentation (by image processing techniques) at the input does not ensure efficient performance across all other MRI datasets, thus, it may lead to lack of generalization. In contrast, our proposed method generates augmented features automatically in the middle of the CNN model. As a consequence, this data augmentation becomes less reliant on the specific statistics of a given dataset, thereby enhancing the generalization ability of the model. 
\end{enumerate}

\subsection{Mathematical Analysis of parallel concatenation block (pcb)}
In this research paper, we have also presented a \textit{lemma}, in order to support our proposed framework. This \textit{lemma} and its mathematical analysis further strengthen our proposed method.

\vspace{0.1cm}
\textit{Lemma: Let's consider two different basic units of the proposed model, (I) Model $c_1$ (shown in fig.2a) in which $z_1$ and $z_2$ convolutional blocks are connected in parallel. (II) Model $c_2$ (shown in fig.2b), in which no parallel concatenation is deployed. If the parallel concatenation block $z_1$ in $c_1$ extracts distinct features with respect to the features extracted in $z_2$, then the model $c_1$ generalizes better than the sequential model $c_2$, especially for the minor classes. As a consequence, the CNN model can automatically mitigate the class imbalance issue to a certain extent. }

\textit{Analysis:} Direct proof of this \textit{Lemma} is not possible. Previously our research group proposed similar kind of lemma in [42], however, mathematical analysis was not provided. In this paper, we have tried to analyze this \textit{lemma} under certain circumstances. Let's assume, the efficacy in major class, that is, $p_{k,H}$ in terms of probability, is already higher (closer to 1) in $c_2$ framework, however, $p_{j,m}$ in minor class hinders with poor efficacy because it has less number of data instances. This kind of circumstance is very common in class imbalance problem. 

\begin{equation}
		That \hspace{0.1cm} means, \hspace{0.3cm} p_{j,m}\ll p_{k,H} \hspace{0.1cm} for \hspace{0.1cm} c_2
\end{equation}
Here, $p_{j,m}$ is the probability of $j^{th}$ sample from minor class ($m$) is correctly classified, $p_{k,H}$ is the probability of $k^{th}$ sample from major class ($H$) is correctly classified.

Now we have to show that after incorporating block $z_1$ parallelly with $z_2$ in $c_1$ model, 

\begin{equation}
p_{j,m}\approx p_{k,H} 
\end{equation}

Then only class imbalance problem will be alleviated according to this \textit{lemma}.

Now, in order to analyze these probability values, we need to further investigate more deeper, that is analyzing layer-wise output in both of the frameworks $c_1$ and $c_2$.

$p_{j,m}$ in system $c_2$ (from Fig.2b) can be estimated as the following formula, 
\begin{equation}
p_{j,m,c2}=D_{c_2}(f^{GAP}(f^{z_2}(f^{z_1}(x_m))))
\end{equation}
Here in equation (6), $x_m$ is the input tensor coming from $z$ block (for minor class), $f^{z_1}$ and $f^{z_2}$ are the features extracted in $z_1$ and $z_2$ convolutional blocks respectively, $f^{GAP}$ is a function for the GAP layer, $D_{c_2}$ is the Single Layer Perceptron (SLP) and the final output coming out from this $D_{c_2}$ will be in between $0$ to $1$, due to employing Softmax activation function at the output layer. Thus, $p_{j,m,c2}$ is the final probability of classifying a sample from minor class `$m$' correctly.  

Similarly $p_{j,m}$ in the system $c_1$ can be estimated as the following formula, 
\begin{equation}
p_{j,m,c1}=D_{c_1}(f^{GAP}(f^{z_2}(x_m)+f^{z_1}(x_m)))
\end{equation}
Here, in equation (7), $x_m$ is the input tensor coming from $z$ block for minor class, $f^{z_1}$ are the augmented features extracted by the $z_1$ block, after passing it to the Avg-2Max pooling. The meaning of other components in equation (7) is exactly the same as in equation (6).

Comparing equations (6) and (7), we can conclude that features coming into GAP for the system $c_1$, have more information than that of the system $c_2$, if and only if 
\begin{equation}
    f^{z_1}\neq f^{z_2}
\end{equation}

Equation (8) is true for our proposed model. Due to incorporating Avg-2Max pooling layer, the overall statistics of the extracted features differ significantly in ${z_1}$ and ${z_2}$ blocks. From the equation (7) it can be observed that, probability $p_{j,m,c1}$ we have obtained after adding both conventional features $f^{z_2}$ and augmented features $f^{z_1}$.

On the other hand, $p_{j,m,c2}$ is computed by only conventional features $f^{z_1}$ which might be overwritten by another conventional feature $f^{z_2}$ (because of the series connection). 

Hence, is likely that Entropy ($E$) of features coming from system $c_1$ is greater than that of system $c_2$. 
\begin{equation}
   E(f^{z_2}(x_m)+f^{z_1}(x_m))> E(f^{z_2}(f^{z_1}(x_m)))
\end{equation}
Hence, there will be more number of neurons and more valuable features in $f^{GAP}$ for system $c_1$ than that of $c_2$. Consequently, system $c_1$ can automatically push the probability value $p_{j,m}$ (of classification) to a higher value, especially in minor class.

\vspace{0.1cm}
For major class also the efficacy (or, class probability) $p_{k,H}$ in the system $c_1$, can be estimated as the following formula,
\begin{equation}
p_{k,H,c1}=D_{c_1}(f^{GAP}(f^{z_2}(x_H)+f^{z_1}(x_H)))
\end{equation}
Here, $x_H$ are the input tensors coming from the $z$ block, for the major class, $p_{k,H}$ is the probability of classifying $k^{th}$ sample correctly into major class $H$. The other components in equation (10), are exactly same as equation (7).   
From equation (10) it is evident that, parallel concatenation block (pcb) will increase the augmented feature maps for the major classes as well. However, this was already assumed (at the starting of the analysis) that $c_2$ model has higher efficacy for the major class, which means, $p_{k,H}\approx1$. So there is a possibility that for major class, there may not be a notable improvement. Because in that case, the CNN model may have already gone to its saturation level. Hence, in this case, we can conclude that after incorporating parallel concatenation block $z_1$, only classification efficacy for minor class, that is $p_{j,m}$, will be boosted so that, it may be somehow comparable with the value of $p_{k,H}$, that is, classification efficacy of the major class. 
\begin{equation}
p_{j,m}\approx p_{k,H} \hspace{0.2cm} (Hence \hspace{0.1cm} proved)
\end{equation}

\subsection{Training Specifications:}
All of the CNN models have been built using Keras sequential API. Tesla P100 GPU was provided by Google Colab Pro service. 
	The following training specifications are followed for overall all the existing CNN models. 
	
\begin{enumerate}
		\item All three MRI datasets are split into training, testing and validation with a ratio of 60\%-20\%-20\%. This splitting is done in a stratified way, which is feasible for class imbalance problem.
		\item Adams-optimizer is employed as the preferred choice of optimizer for all the models. 
		\item A batch size of 8 is utilized while training all the CNN models. 
        \item All the images are resized into 224 x 224 before feeding them into all the CNN models. 
        \item In all pre-trained CNN models, no FC layer (or, dense layer) is taken into account, in order to avoid over-fitting problem. This is also feasible for fair comparison with the proposed model.  
        \item This is to clarify that these Fibonacci blocks are proposed in a flexible way. One may choose 6, 7 or 8 blocks based on the size and complexity of the dataset. For instance, we have chosen Fibonacci Net with 6 blocks for small 4-class dataset, and employed 7 blocks for both 15 and 44-class datasets. Whereas, the main novelty lies in $3^{rd}$ and $4^{th}$ block, due to these skip connections.  
        \item We have also incorporated adaptive learning rate (alr) in the proposed framework, in which for the first 13 epochs we maintain fixed lr $1e^{-4}$, thereafter, it will be decaying by a factor of 0.9 upto 25 epochs. These numbers are chosen empirically. This adaptive learning is part of our model novelty, and only employed in the proposed model.
        \item Early stopping is not employed in any of the model implementations, since it has been observed that early stopping often stops the training too early for all three MRI datasets. 
        \item No pre-processing method, like data-augmentation or image processing technique is deployed in any of the experiments. 	
\end{enumerate}

\begin{table*}[tb]
\begin{center}
\caption{Comparisons of existing CNN / ViT models with the proposed ``Fibonacci Net'' on testing (Weighted Average)}
\label{tab:my-table}
\resizebox{2.1\columnwidth}{!}
{
\begin{tabular}{|c|ccc|ccc|ccc|c|}
\hline
\multirow{3}{*}{\begin{tabular}[c]{@{}c@{}}Models\textbackslash\\ Methods\end{tabular}}          & \multicolumn{3}{c|}{4-class dataset}                                                                                                & \multicolumn{3}{c|}{15-class dataset}                                                                                                        & \multicolumn{3}{c|}{44-class dataset}                                                                                              & \multirow{3}{*}{\begin{tabular}[c]{@{}c@{}}No of \\ params\\ (in Lakhs)\end{tabular}} \\ \cline{2-10}
                                                                                                 & \multicolumn{1}{c|}{Accuracy}       & \multicolumn{1}{c|}{F1-score}       & \begin{tabular}[c]{@{}c@{}}secs/ \\ epochs\end{tabular} & \multicolumn{1}{l|}{Accurcay}       & \multicolumn{1}{c|}{F1-score}       & \textbf{\begin{tabular}[c]{@{}c@{}}secs/ \\ epochs\end{tabular}} & \multicolumn{1}{l|}{Accuracy}       & \multicolumn{1}{l|}{F1-score}       & \begin{tabular}[c]{@{}c@{}}secs/\\ epochs\end{tabular} &                                                                                       \\ \hline
\begin{tabular}[c]{@{}c@{}}VGG-16 (pretrained)\end{tabular}                             & \multicolumn{1}{c|}{0.283}          & \multicolumn{1}{c|}{0.125}          & 87                                                      & \multicolumn{1}{c|}{0.103}          & \multicolumn{1}{c|}{0.103}          & 158                                                              & \multicolumn{1}{c|}{0.031}          & \multicolumn{1}{c|}{0.031}          & 290                                                    & 158                                                                                   \\ \hline
\begin{tabular}[c]{@{}c@{}}ResNet-50 (pretrained)\end{tabular}                          & \multicolumn{1}{c|}{0.683}          & \multicolumn{1}{c|}{0.683}          & 26                                                      & \multicolumn{1}{c|}{0.896}          & \multicolumn{1}{c|}{0.901}          & 28                                                               & \multicolumn{1}{c|}{0.596}          & \multicolumn{1}{c|}{0.603}          & 73                                                     & 280                                                                                   \\ \hline
\begin{tabular}[c]{@{}c@{}}Inception-V3 (pretrained)\end{tabular}                       & \multicolumn{1}{c|}{0.858}          & \multicolumn{1}{c|}{0.857}          & 11                                                      & \multicolumn{1}{c|}{0.905}          & \multicolumn{1}{c|}{0.910}          & 128                                                              & \multicolumn{1}{c|}{0.672}          & \multicolumn{1}{c|}{0.678}          & 105                                                    & 275                                                                                   \\ \hline
\begin{tabular}[c]{@{}c@{}}DenseNet-121 (pretrained)\end{tabular}                       & \multicolumn{1}{c|}{0.946}          & \multicolumn{1}{c|}{0.946}          & 39                                                      & \multicolumn{1}{c|}{0.837}          & \multicolumn{1}{c|}{0.836}          & 190                                                              & \multicolumn{1}{c|}{0.758}          & \multicolumn{1}{c|}{0.754}          & 90                                                     & 92                                                                                    \\ \hline
\begin{tabular}[c]{@{}c@{}}MobileNet-V2 (pretrained)\end{tabular}                       & \multicolumn{1}{c|}{0.886}          & \multicolumn{1}{c|}{0.886}          & 12                                                      & \multicolumn{1}{c|}{0.889}          & \multicolumn{1}{c|}{0.890}          & 8.9                                                              & \multicolumn{1}{c|}{0.732}          & \multicolumn{1}{c|}{0.744}          & 3.4                                                    & 32                                                                                    \\ \hline
\begin{tabular}[c]{@{}c@{}}DCNN by [12] \\(trained from scratch)\end{tabular}     & \multicolumn{1}{c|}{0.890}          & \multicolumn{1}{c|}{0.888}          & 31                                                      & \multicolumn{1}{c|}{0.801}          & \multicolumn{1}{c|}{0.801}          & 93                                                               & \multicolumn{1}{c|}{0.736}          & \multicolumn{1}{c|}{0.736}          & 29                                                     & \textbf{4.14}                                                                         \\ \hline
\begin{tabular}[c]{@{}c@{}}Brain MR-Net$+$CBAM [26]\\ (trained from scratch)\end{tabular}             & \multicolumn{1}{c|}{0.592}          & \multicolumn{1}{c|}{0.594}          & 25                                                      & \multicolumn{1}{c|}{0.702}          & \multicolumn{1}{c|}{0.692}          & 17                                                               & \multicolumn{1}{c|}{0.662}          & \multicolumn{1}{c|}{0.642}          &  2                                                      & 6.16                                                                                  \\ \hline
\begin{tabular}[c]{@{}c@{}}Pooling based ViT\\ (PiT) [21], pre-trained\end{tabular} & \multicolumn{1}{c|}{0.894}          & \multicolumn{1}{c|}{0.894}          & 21                                                      & \multicolumn{1}{c|}{0.739}          & \multicolumn{1}{c|}{0.736}          & 31                                                               & \multicolumn{1}{c|}{0.722}          & \multicolumn{1}{c|}{0.722}          & 31                                                     & 46                                                                                    \\ \hline
\begin{tabular}[c]{@{}c@{}}ConvNext-V2 [46]\\ (pre-trained)\end{tabular} & \multicolumn{1}{c|}{0.318}          & \multicolumn{1}{c|}{0.001}          & 21                                                      & \multicolumn{1}{c|}{0.948}          & \multicolumn{1}{c|}{0.949}          & 33                                                               & \multicolumn{1}{c|}{0.952}          & \multicolumn{1}{c|}{0.954}          & 33                                                     & 494                                                                                    \\ \hline
\begin{tabular}[c]{@{}c@{}} {\textbf{Proposed Fibonacci-Net}} \\(trained from scratch)\end{tabular}                     & \multicolumn{1}{c|}{\textbf{0.950}} & \multicolumn{1}{c|}{\textbf{0.950}} & 7                                             & \multicolumn{1}{c|}{\textbf{0.955}} & \multicolumn{1}{c|}{\textbf{0.953}} & 13                                                      & \multicolumn{1}{c|}{\textbf{0.962}} & \multicolumn{1}{c|}{\textbf{0.965}} & 12                                            & 13.95                                                                                 \\ \hline
\end{tabular}
}
\end{center}
\end{table*}

\section{Results and Analysis}
Results and Analysis section can be summarized into main two parts: (a) Comparison of the proposed framework with several state-of-the-art models, (b) Ablation study of the proposed framework, (c) Validity checking of the model with the help of Explainable AI.
As it is a class imbalance problem, precision, recall, F1 score, and AUC [45] are also taken into account along with accuracy, as quality evaluation metrics. Moreover, all the codes with graphs, metrics, classification reports and confusion matrices are shared in a GitHub link \textbf{(https://anonymous.4open.science/r/Brain-Tumor-Detection/)}.

\begin{figure*}[h]
		\centering
\includegraphics[width=18.0cm,height=3.8cm]{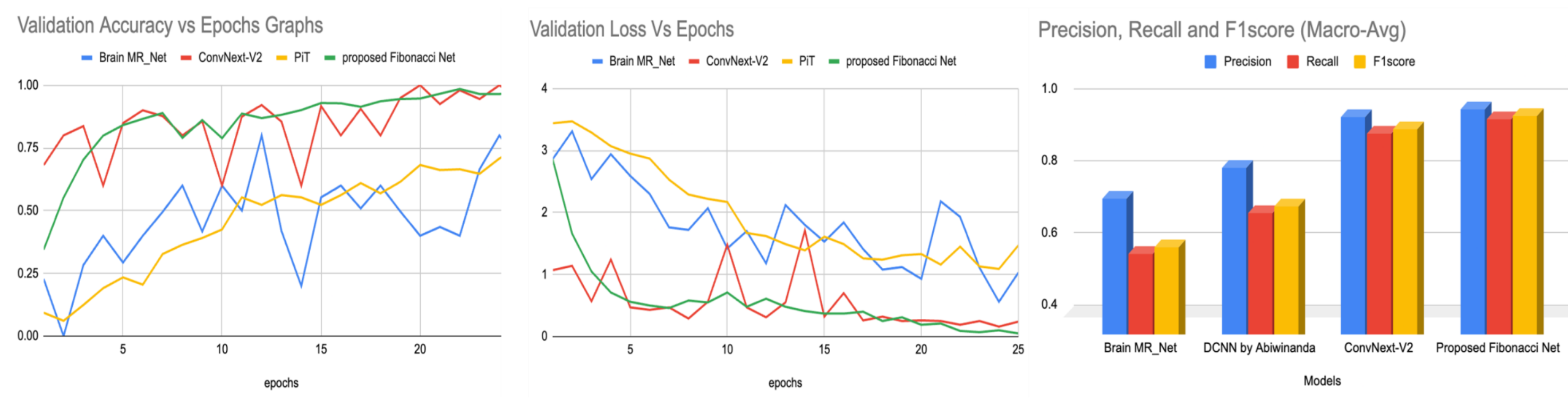}
		\caption{$1^{st}$ and $2^{nd}$ graphs represent validation accuracy and validation loss (vs number of epochs) for 44-class dataset, $3^{rd}$ graph shows Macro-average of Precision, Recall and F1 score for several models on 44-class dataset}
	\end{figure*}

\subsection{Comparison of proposed framework with state-of-the-art methods}
We have implemented numerous standard pre-trained CNN models VGG-16, ResNet-50, Inception-V3, DenseNet-121, MobileNet etc. on all three MRI datasets. All these models are 100\% fine-tuned (transfer learning done) from ImageNet dataset. Additionally, we have also implemented two existing CNN architectures, namely DCNN by Abiwinanda et al. [12] and Brain MR-Net + CBAM [26], both of which were trained from scratch and specifically proposed for brain tumor detection from MRI scans. Additionally, we have performed two recent trends models, the Pooling-based Vision Transformer (PiT) model [21] and Convnext-V2 [46] on all three MRI datasets and compared their efficacy with the same of the proposed Fibonacci Net. Experimental results in Table-I reveal that the proposed model ``Fibonacci Net'' (trained from scratch) has consistently outperformed all the pre-trained models and existing trained-from-scratch models, across all three MRI datasets. 

\begin{table*}[tb]
\begin{center}
\caption{Ablation Studies of the proposed ``Fibonacci-Net'' on all three MRI Datasets}
\label{tab:my-table}
\resizebox{2.0\columnwidth}{!}
{
\begin{tabular}{|c|cccl|cccl|cccl|}
\hline
\multirow{3}{*}{\begin{tabular}[c]{@{}c@{}}Fibonacci Net\\ with several \\ combinations\end{tabular}} & \multicolumn{4}{c|}{4-class dataset}                                                                                                                                                                         & \multicolumn{4}{c|}{15-class dataset}                                                                                                                                            & \multicolumn{4}{c|}{44-class dataset}                                                                                                                                                                        \\ \cline{2-13} 
                                                                                                      & \multicolumn{1}{c|}{Accuracy}       & \multicolumn{1}{c|}{\begin{tabular}[c]{@{}c@{}}F1-\\ score\end{tabular}} & \multicolumn{1}{c|}{\begin{tabular}[c]{@{}c@{}}Preci-\\ sion\end{tabular}} & Recall         & \multicolumn{1}{l|}{Accurcay}       & \multicolumn{1}{c|}{F1-score}       & \multicolumn{1}{c|}{\textbf{\begin{tabular}[c]{@{}c@{}}Preci-\\ sion\end{tabular}}} & Recall         & \multicolumn{1}{l|}{Accuracy}       & \multicolumn{1}{l|}{\begin{tabular}[c]{@{}l@{}}F1-\\ score\end{tabular}} & \multicolumn{1}{c|}{\begin{tabular}[c]{@{}c@{}}Preci-\\ sion\end{tabular}} & Recall         \\ \hline
\begin{tabular}[c]{@{}c@{}}Base model (No Concat)\end{tabular}                                     & \multicolumn{1}{c|}{0.915}          & \multicolumn{1}{c|}{0.915}                                               & \multicolumn{1}{l|}{0.915}                                                 & 0.915          & \multicolumn{1}{c|}{0.941}          & \multicolumn{1}{c|}{0.942}          & \multicolumn{1}{c|}{0.953}                                                          & 0.932          & \multicolumn{1}{c|}{0.942}          & \multicolumn{1}{c|}{0.944}                                               & \multicolumn{1}{c|}{0.958}                                                 & 0.934          \\ \hline
Base model$+$2-4 concat                                                                                            & \multicolumn{1}{c|}{0.924}          & \multicolumn{1}{c|}{0.925}                                               & \multicolumn{1}{c|}{0.926}                                                 & 0.924          & \multicolumn{1}{c|}{0.942}          & \multicolumn{1}{c|}{0.939}          & \multicolumn{1}{c|}{0.945}                                                          & 0.935          & \multicolumn{1}{c|}{0.946}          & \multicolumn{1}{c|}{0.946}                                               & \multicolumn{1}{c|}{0.951}                                                 & 0.942          \\ \hline
\begin{tabular}[c]{@{}c@{}}Base model$+$3-5 concat \\ (16 filters)\end{tabular}                                    & \multicolumn{1}{c|}{0.931}          & \multicolumn{1}{c|}{0.931}                                               & \multicolumn{1}{c|}{0.931}                                                 & 0.931          & \multicolumn{1}{c|}{0.935}          & \multicolumn{1}{c|}{0.937}          & \multicolumn{1}{c|}{0.946}                                                          & 0.928          & \multicolumn{1}{c|}{0.949}          & \multicolumn{1}{c|}{0.951}                                               & \multicolumn{1}{c|}{0.955}                                                 & 0.947          \\ \hline
\begin{tabular}[c]{@{}c@{}}Base model$+$3-5 concat \\ (24 filters)\end{tabular}                                    & \multicolumn{1}{c|}{0.938}          & \multicolumn{1}{c|}{0.937}                                               & \multicolumn{1}{c|}{0.938}                                                 & 0.936          & \multicolumn{1}{c|}{0.941}          & \multicolumn{1}{c|}{0.942}          & \multicolumn{1}{c|}{0.947}                                                          & 0.938          & \multicolumn{1}{c|}{0.952}          & \multicolumn{1}{c|}{0.951}                                               & \multicolumn{1}{c|}{0.963}                                                 & 0.941          \\ \hline
\begin{tabular}[c]{@{}c@{}}Base model$+$2-4 concat\\ $+$3-5 concat (24 filters)\end{tabular}                        & \multicolumn{1}{c|}{\textbf{0.950}} & \multicolumn{1}{c|}{\textbf{0.950}}                                      & \multicolumn{1}{c|}{\textbf{0.950}}                                        & \textbf{0.950} & \multicolumn{1}{c|}{\textbf{0.955}} & \multicolumn{1}{c|}{\textbf{0.953}} & \multicolumn{1}{c|}{\textbf{0.957}}                                                 & \textbf{0.949} & \multicolumn{1}{c|}{\textbf{0.962}} & \multicolumn{1}{c|}{\textbf{0.965}}                                      & \multicolumn{1}{c|}{\textbf{0.971}}                                        & \textbf{0.959} \\ \hline
\end{tabular}
}
\end{center}
\end{table*}

From Table-1 it is evident that, overall, all the pre-trained CNN models exhibited descent performance on both the 4-class and 15-class datasets except VGG-16. It is observed that VGG-16 model generally struggles with the class imbalance issue [29] due to its limited feature extraction capabilities [23]. Inception-V3 [41], DenseNet-121 [47], and MobileNet-V2 [39] achieved accuracy of 85.8\%, 94.6\%, and 88.6\% respectively on 4-class dataset. These models performed well primarily due to their lightweight architecture and extensive use of parallel concatenation blocks. However, the 44-class dataset posed a significant challenge due to the higher number of classes, moreover, class imbalance issue was very substantial here. Many of the classes only had 17 or 28 images. Consequently, overfitting was inevitable and these models could not generalize well on this 44-class MRI dataset. Moreover, from Table-1 it can be observed that trained-from-scratch model ``Brain-MR-Net'' could not perform well throughout all three MRI datasets. Possibly due to heavy usage of interpolation there was huge data loss by their model. Furthermore, because of having very few trainable parameters model goes under-fitting for all three MRI datasets. The recent trend model Pooling-based ViT (PiT) [21] demonstrated reasonable efficacy for the 4-class dataset, however, it underperformed on the 15-class and 44-class datasets. We have found that Vision transformer models mostly struggled with scenarios involving severe class imbalance. On the contrary, the recent trend CNN model ConvNext Version-2 [46] has suffered from poor accuracy and F1 score on the 4-class (small) dataset, however, it performed exceptionally well on both the 15-class and 44-class MRI datasets. Due to having many skip connections, ConvNext-V2 model overcomes the class imbalance problem to some extent (according to our theory in III-C). In fact, ConvNext-V2 achieved the second-highest accuracy on the 15-class and 44-class datasets, surpassed only by our proposed Fibonacci Net by 1\%. Overall, all these existing models have worked well on a specific MRI dataset, nevertheless, they do not ensure effective generalization across diverse (all 3 MRI) datasets.

\begin{figure*}[h]
		\centering
\includegraphics[width=11.0cm,height=9.6cm]{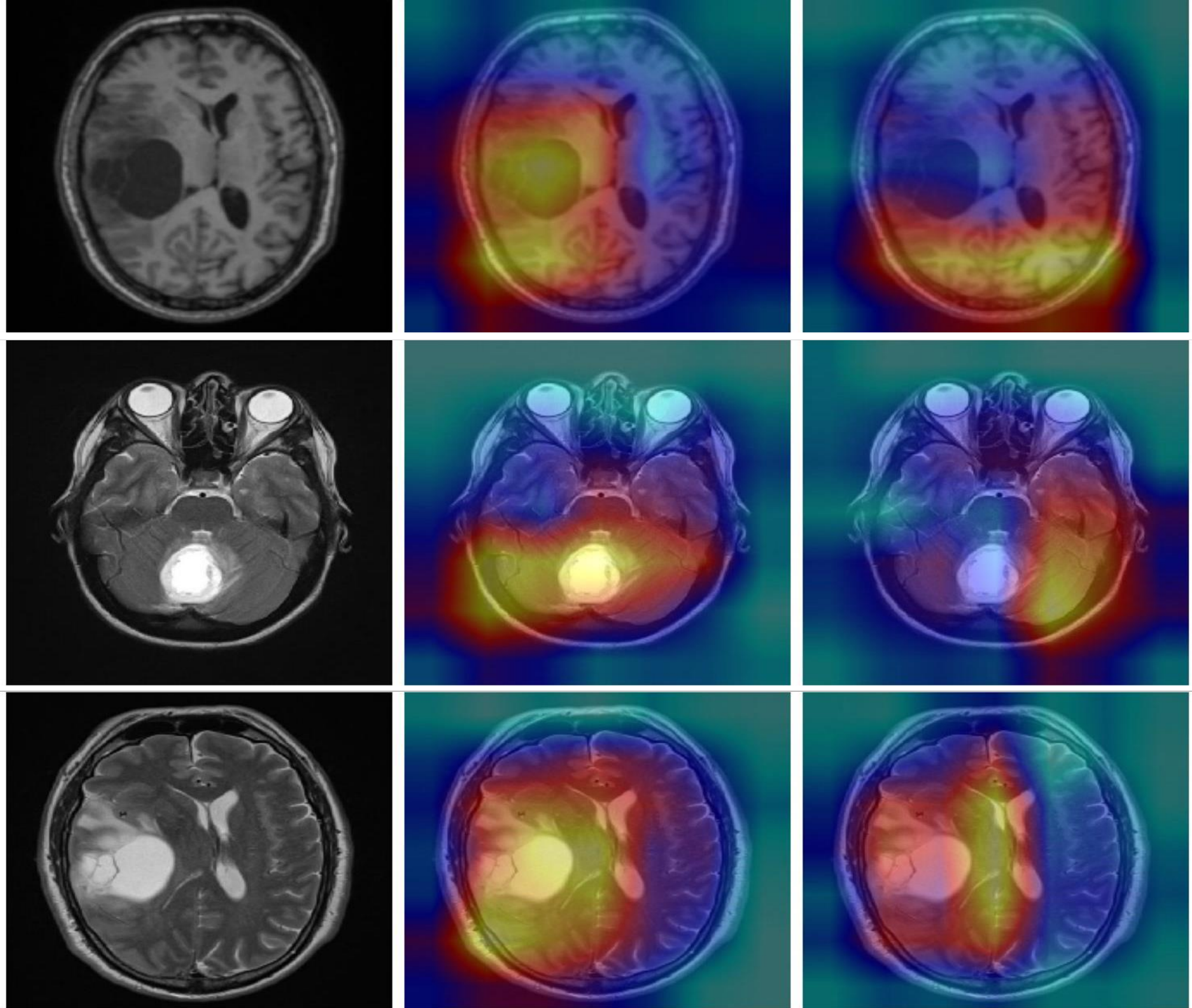}
		\caption{\textbf{Validity checking by Explainable AI:} First column represents the original image, $2^{nd}$ column represents GradCam plot by Proposed ``Fibonacci Net with pcb'', $3^{rd}$ column represents GradCam plot by Proposed ``Fibonacci Net without pcb''}
	\end{figure*}
Table-I also indicates that our proposed model is the only framework that performed efficiently and consistently across all three MRI datasets. It achieved the highest accuracy 95\%, 95.5\% and 96.2\% on 4-class, 15-class and 44-class datasets respectively. Moreover, secs/ epochs is also presented in Table-I, which indicates the model complexity during training. For instance, it can be observed that it took only (average of) 7 secs per epoch during training the proposed model on 4-class MRI dataset, which is considerably lesser than that of other models. Hence, this proves that the proposed model complexity is quite lower than the other existing methods. A graph representation of macro averages of precision, recall, and F1-score is also presented in Fig.4 last column. This macro average is the average across all classes. Thus, often this Macro average reflects the class imbalance issue well. The Fig.4 reveals that for the 44-class the Macro-average metrics of the proposed model outperformed all other existing models as well. Only the performance of ConvNext-V2 was comparable with the proposed model. Hence, from Fig.4 and Table-I, it can be concluded that the proposed framework substantially mitigated the issue of class imbalance and overall it generalized well across all 3 MRI datasets.

Along with Table-I, we have also presented a comparative graph analysis, in Fig. 4, implemented on highly imbalanced `44-class MRI' dataset. We deliberately selected only three recent trend models for comparison, otherwise, the graphs could become too complex to interpret. From the Fig.4, it is evident that both Brain MR-Net and Pooling based Vision Transformer (PiT) could not achieve validation accuracy exceeding 80\% over the course of 25 epochs, possibly due to large class imbalance present in the dataset. In contrast, the ConvNext model surpassed the performance of these two existing models and is the only one comparable to our proposed model. However, the graphs of both validation accuracy vs epochs and loss vs epochs, for the ConvNext model have shown significant fluctuations throughout 25 epochs. Thus, the performance of the ConvNext-V2 model is not stable. On the other hand, our proposed Fibonnacci-Net has shown much smoother and stable validation graphs (shown in green color in Fig.4) than any other existing methods, thereby indicating that the proposed model generalized very well.

% \usepackage{multirow}

\subsection{Ablation Studies of the proposed Framework}
We have presented an ablation study (in brief) on the proposed ``Fibonacci Net'' in Table-II. We experimented with various combinations, such as 2-4 concatenation, 3-5 concatenation, and a combination of both 2-4 concatenation + 3-5 concatenation etc. as skip connections, on top of the ``Fibonacci Net" base model. As depicted in Table II, only the framework with the combination of 2-4 concat + 3-5 concat (with 24 filters) proved to be efficient and consistent across all three MRI datasets. The individual experiments with only 2-4 concat or only 3-5 concat (24 filters) showed slight improvements in the model’s performance. These improvements were particularly noticeable for the 4-class and 44-class datasets. However, for the 15-class dataset, it might be misconstrued that utilizing 3-5 or, 2-4 concat does not lead to any improvement in accuracy or F1 score. This is apparent from Table-II that the difference between precision and recall gets reduced by 2-4 or 3-5 concat, which is a possible indication that class imbalance is getting eliminated. Hence, we opted to use both these skip connections simultaneously on the base model for all 3 MRI datasets. This particular combination was found empirically, after lots of combinations tried. It is quite evident from Table-II that, this particular combination has improved the model accuracies from 91.5\% to 95\% for 4-class dataset. Similarly for 15-class dataset and 44-class dataset these improvements are approximately 1.5\% and 2\% respectively. Hence, it can be concluded that these skip connections helped the proposed framework to boost the accuracy significantly. This subsection of ``results and analysis'' also supports our proposed lemma and mathematical analysis provided in section III-{C}.

\subsection{Validity checking of the proposed theory by Explainable AI}
To validate our proposed theory, we have also utilized explainable AI technique [48] in this paper. A Grad-CAM heat map visualization [48] is presented in Fig. 5. Here, all three original images in the $1^{st}$ column are taken from 44-class MRI dataset. The image in the  $1^{st}$ row, $2^{nd}$ row and $3^{rd}$ row (of Fig.5) are taken from T1, T1C+, and T2 modes respectively. As evident from Fig. 5, in the second column, our proposed model ``Fibonacci Net'' focuses more intensely on the tumor regions (or, suspicious regions) for all the modes, after incorporating pcb blocks. In the $3^{rd}$ column it is observed that the proposed Fibonacci Net without pcb block could not align directly on the suspicious (tumor) regions for the first two images. Nevertheless, for the third image, it aligns on the tumor region, to some extent. This indicates that by integrating two skip connections or pcb blocks (depicted in Fig.1), the model now able to generate more essential and variety of (augmented) features heavily in the middle of the model. This perhaps improves the attention of the model during training. As a consequence, it significantly enhances the generalization ability of the proposed model, thereby enabling the model to distinguish among various classes of tumors very efficiently.

\section{Conclusion}
A ``Fibonacci Net'' along with pcb connections was proposed for brain tumor classification task from MRI datasets. A prevalent issue of class imbalance across diverse MRI datasets was first identified. The proposed framework was a very lightweight model, having very few trainable parameters (1.3 million). Moreover, the pcb blocks in the proposed framework worked like an automatic feature augmenter which generated diverse set of (augmented) features in the middle of the CNN model, thereby improving the generalization ability of the model. A novel pooling technique ``Avg-2Max Pooling'' was deployed in these pcb blocks for the first time, resulting in altering the overall statistics of the augmented feature maps from those of the original feature maps. Hence, this pcb blocks worked like an efficient feature augmenter, thereby alleviating the class imbalance issues to a certain extent. A lemma and mathematical analysis was provided which strengthened our proposed theory. Experimental results in section IV showed that our proposed framework surpassed the efficacy of a recent trend Vision Transformer, and latest ConvNext-V2 model. Furthermore, a GradCam heat map visualization validated our proposed theory of pcb connections.



\end{document}